\documentclass[aps,pra,twocolumn,showpacs,superscriptaddress,twoside,floatfix]{revtex4-1}
\pdfoutput=1

\usepackage{siunitx}
\usepackage{amsmath}
\usepackage{amsfonts}
\usepackage{hyperref}
\usepackage[normalem]{ulem}
\usepackage{graphicx,xcolor}

\usepackage[english]{babel}

\newcommand{\pdd}[2]{\frac{\partial #1}{\partial #2}}
\newcommand{\pdds}[2]{\frac{\partial^2 #1}{\partial #2^2}}
\newcommand{\spdd}[1]{\pdd{}{#1}}
\newcommand{\spdds}[1]{\pdds{}{#1}}

\newcommand {\ft}[1]{\mathcal{FT}\left(#1\right)}

\newcommand{\dd}[2]{\frac{d #1}{d #2}}
\newcommand{\dds}[2]{\frac{d^2 #1}{d#2^2}}

\newcommand{\eq}[1]{Eq.~\eqref{eq:#1}}
\newcommand{\fig}[1]{Fig.~\ref{fig:#1}}

\newcommand{\solidrule}[1][5mm]{\rule[0.5ex]{#1}{1pt}}
\newcommand\thickrule[1][5mm]{\rule[0.4ex]{#1}{2pt}}

\newcommand\thickdashedrule[1][1mm]{\protect{\mbox{%
		  \thickrule[#1]\hspace{#1}\thickrule[#1]\hspace{#1}\thickrule[#1]}}}

\definecolor{c4}{rgb}{0.2,0.6274,0.1725}

\newcommand{\AnalyticR}{dashed black, \textcolor{black}{\thickdashedrule}}
\newcommand{\AnalyticT}{dashed black, \textcolor{black}{\thickdashedrule}}
\newcommand{\Rstyle}{orange, \textcolor{orange}{\thickrule}}
\newcommand{\Tstyle}{green, \textcolor{c4}{\thickrule}}
\newcommand{\Lstyle}{black, \textcolor{black}{\solidrule}}

\begin{document}

\title{Coupled mode approach to square gradient Bragg reflection resonances \\in corrugated
dielectric waveguides}

\author{Otto Dietz}
\email{otto.dietz@physik.hu-berlin.de}
\author{G\"{u}nter Kewes}
\author{Oliver Neitzke}
\author{Oliver Benson}
\affiliation{Nano-Optics Group, Institut f\"{u}r Physik, Humboldt-Universit\"{a}t zu Berlin, Germany}

\date{\today}

\begin{abstract}
		  We demonstrate the appearance of unexpected reflection resonances in
		  corrugated dielectric waveguides. These are due to the
		  curvature of the boundary. The effect is as strong as the ordinary Bragg
		  resonances, and reduces the transmission through our
		  waveguide by 20\%. It is thus of high relevance for the design of
		  optimized waveguiding structures. We validate our analytical predictions based on
		  coupled mode theory by a comparison to numerical simulations.
\end{abstract}



\maketitle


\section{Introduction}
Light scattering is the key mechanism to tailor the properties of passive
waveguiding structures. Ordered and disordered boundaries determine the
transmission, the reflection, and the radiation losses. In highly integrated
structures these processes can be controlled through designed lithography, 
to modify light propagation at will.
For this reason, a thorough understanding of scattering processes is
mandatory.  One powerful tool for understanding the light propagation and
scattering is the concept of Bragg scattering. Given only the periodicity of a
waveguiding structure, it predicts reflection resonances in a straightforward
manner. 
Therefore, light propagation through dielectric waveguides, has been intensively
investigated concerning Bragg scattering
\cite{tan_chip-scale_2008,lavdas_wavelength_2014,winick_design_1990,kogelnik_coupled-wave_1972,flanders_grating_1974,kewes_key_2015}.

In this work we show that care has to be taken when applying Bragg scattering
analysis in a too simplified way. Surprisingly, rather strong reflection
resonances can be overseen.

In a waveguide with periodic boundaries  the $m$-th order Bragg reflection is in
general expected for the
wavelength $\lambda$ that (for perpendicular incident) fulfills
\begin{equation}
		  2d=m\lambda
		  \label{eq:generalbragg}
\end{equation}
where $d$ is the length of the periodicity, e.g., the lattice spacing.

\begin{figure}
					 \includegraphics[width=0.5\textwidth]{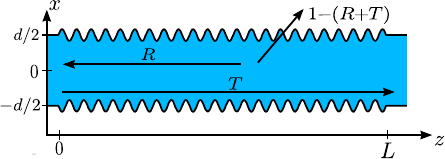}
		  \caption{Model of corrugated waveguide. Transmittance $T$,
					 reflectivity $R$, and losses $1-(R+T)$. The (mean) width of the
					 waveguide is $d=450$nm, the length $L=10d=4.5\mu m$. The refractive
					 index of the inner (outer) material is $n=2$
					 ($n=1$). The
					 wavelength of the boundary oscillation is $\Lambda=200$nm, and its amplitude
					 is $A=37.5$nm ($\sigma=37.5\textrm{nm}/\sqrt{2}$). 
					\label{fig:setup}}
\end{figure}

The appearance of multiple order $m$ seems obvious, because reflection takes
place, whenever the backscattered wave interferes constructively with the
incident wave. This is always fulfilled for wavelength increments of $2d$.
Surprisingly this simple picture of multiple order is incorrect for some systems. There are
periodic systems where only a single Bragg reflection ($m=1$) exists. 
For example, waveguides with infinite sinusoidal boundaries (similar to the
finite waveguide sketched in \fig{setup}). The Fourier series of the
sinusoidal boundary consists of only a single (positive) coefficient.
Therefore these systems are believed to show a single Bragg reflection only \cite{yariv_optical_2003}.

Here we perform a more rigorous analysis and show that this is in fact not true.  
It turns out that corrugated waveguides show an additional Bragg reflection
which is not expected from previous studies on individual periodic systems
\cite{kogelnik_2._1975,yariv_optical_2003}. 
Since the simple Bragg picture in \eq{generalbragg} often serves as starting point for numerical design of
optical components \cite{winick_design_1990,kewes_key_2015}, it is vital to know
which resonances can exist in principle.

\fig{setup} shows a typical finite waveguide which we study in our
analysis more explicitly as an example.  We apply
a technique from statistical boundary roughness analysis \cite{izrailev_manifestation_2006} 
to single dielectric waveguides. 
Within a coupled mode approach we are able to \textit{generalize} findings for ensembles
of systems  to single systems under drastically \textit{relaxed}
assumptions. 
The previous theoretical findings become a special limiting case in
our framework.

The evidence of an additional Bragg resonance presented here,
is important for many systems in a variety of communities, where corrugated
waveguides are employed in very different applications, such as group-velocity control
\cite{tan_chip-scale_2008}, phase-matching in nonlinear materials \cite{lavdas_wavelength_2014}, 
distributed feedback laser \cite{kogelnik_coupled-wave_1972}, optical filtering
\cite{winick_design_1990}, grating couplers \cite{flanders_grating_1974}, and
hybrid atom-photonic systems \cite{yu_nanowire_2014}. 
\section{coupled mode approach}

Coupled mode theory is a powerful and common method \cite{yariv_optical_2003}.
The starting point is the wave
equation for a dielectric waveguide with permeability $\mu$ and dielectric
function $\epsilon(x,y)$, which is weakly perturbed by $\Delta
\varepsilon(x,y,z)$, written as
\begin{equation}
        \left[ \nabla^2 +\omega^2\mu(\epsilon+\Delta\epsilon) \right]\vec
        E(x,y,z)=0.
        \label{eq:diel_weq_perturb}
\end{equation}
The field $\vec E$ of frequency $\omega=k c$ can be constructed from the Eigenfunctions $E_l$ of the
unperturbed waveguide. The contribution $A_l$ of each Eigenmode $E_l$, changes
along the waveguide, due to the dielectric perturbation, such that
\begin{equation} 
        \vec E = \sum_l A_l(z) \vec E_l(x,y)e^{i(\omega t-\beta_mz)}
        \label{eq:fourier_series}
\end{equation}
where $\beta_m$ is the propagation constant of the $m$'th mode. This is, the
component of the wave vector in propagation $z$-direction.
Inserting \eq{fourier_series} into \eq{diel_weq_perturb}
and taking the scalar product with $E^*_k$ yields  (see \cite{yariv_optical_2003} for details)
\begin{equation}
        \dd{}{z}A_k = -i\frac{\beta_k}{|\beta_k|}\sum_l
         C^{(b)}_{kl}A_le^{i(\beta_k-\beta_l)z}
        \label{eq:yariv_6.4-12}
\end{equation}
where the changes in $A_l$ are assumed to be sufficiently ``slow'' to neglect
$\dds{A_l}{z}$ (see \cite{yariv_optical_2003}). The coupling coefficient
\begin{equation}
         C^{(b)}_{kl} =\frac{\omega}{u} \int dx dy \vec E_k \Delta\epsilon \vec E_l
        \label{eq:coupling_coefficient}
\end{equation}
describing the overlap between two modes $k$ and $l$. The coefficient $u$ accounts for different possible choices of the normalization of $\vec E_k$. 
In the following, we will
restrict ourselves to two modes. In this case 
\eq{yariv_6.4-12} is a set of two coupled
equations $(k=1, l=2$ and $k \leftrightarrow l$)
\begin{align}
        \dd{}{z}A_1 &= -i\frac{\beta_1}{|\beta_1|}\sum_2
		  C^{(b)}_{12}A_2e^{i(\beta_1-\beta_2)z}\label{eq:a1}\\
        \dd{}{z}A_2 &= -i\frac{\beta_2}{|\beta_2|}\sum_1
		  C^{(b)}_{21}A_1e^{i(\beta_2-\beta_1)z}\label{eq:a2}
\end{align}
A relevant figure is the reflectivity of a system. Solving these coupled
mode equations yields two solutions $A_1,A_2$. The ratio of these two
solutions at the beginning of the sample ($z=0$), is the ratio of incoming to
backscattered mode, i.e., the reflectivity
\begin{align}
        R &= \left|\frac{A_2(0)}{A_1(0)}\right|^2
        \label{eq:reflection}
\end{align}
%
%
When solving \eq{a1} and \eq{a2} and plugging the solution into
\eq{reflection}, the maximum reflectivity is obtained as 
\begin{equation}
R_{max}=\tanh^2\left(\hat C^{(b)}_{kl}L\right)
		  \label{eq:maxrefl}
\end{equation}
where $\hat C_{kl}$ is the Fourier coefficient in the Fourier series of
$C_{kl}$ (as in \eq{coupling_coefficient_periodic}).

A common simplification is to assume that the dielectric perturbation is periodic in $z$ \cite{flanders_grating_1974,kogelnik_2._1975} 
$$
\Delta
\epsilon(x,y,z)=\sum_m \epsilon_m(x,y) e^{-im\frac{2\pi}{\Lambda}z}
$$
Then the $z$-dependence of $\Delta\epsilon(x,y,z)$ can be
separated from the overlap integral in \eq{coupling_coefficient},
which is only over $x$ and $y$.
\begin{equation}
		  C^{(b)}_{kl} =\frac{\omega}{u} \sum_m e^{-im\frac{2\pi}{\Lambda}z}\int dx dy
		  \vec E_k \epsilon_m(x,y) \vec E_l 
        \label{eq:coupling_coefficient_periodic}
\end{equation}
This means, in particular, that the integral becomes independent of $z$.
In fact we reduced the description of coupling effects to a
stratified waveguide. This is, a waveguide which is composed of
rectangular slices, each with some $\Delta \epsilon$. This is not
surprising, because we constructed the E-Field as contributions of
Eigenfunctions, weighted by $A_l$ in \eq{fourier_series}. 
We assumed that the $(x,y,z)$-dependencies of the E-Field can be
separated into $A_l(z)$ and the Eigenmodes $E_l(x,y)$.
However, if we then try to tackle an arbitrary corrugated waveguide, we have to keep in mind that the only results we can
expect are results for \textit{stratified} waveguides. Common text book
approaches ignore this fact
\cite{yariv_optical_2003,kogelnik_2._1975,ebeling_integrated_2011}. 

\section{coordinate transformation}

Corrugated waveguides feature rich physical effects
which are not present in stratified waveguides. Therefore, as a next step we
will show how to overcome the shortcomings of common coupled mode theory,
without rejecting the entire approach, which, indeed, captivates by its clarity
and simplicity.

As the above problems are of geometrical nature, i.e., restriction
to stratified waveguides, it is reasonable to look
for a geometrical solution. Here we use a straightforward
coordinate transformation to transform the corrugated
boundaries to 
flat boundaries.
%
%
%

This transformation has been used previously to derive the
square gradient scattering mechanism in systems that feature boundaries with
randomized roughness \cite{izrailev_manifestation_2006}. Even though this was the very first
derivation of the square gradient mechanism, the general validity of this
mechanism was so far doubtful for several reasons. At first it is derived for ensembles
of systems that feature peculiar statistical properties. Therefore it is not a
priori clear if the mechanism is only a statistical effect. Previous
experiments \cite{dietz_surface_2012} investigated systems that resembled these
special statistical properties. Furthermore, both theory and experiment have so
far being restricted to hollow waveguides with  perfectly electrically
conducting boundaries.

In the following, we will put the square gradient scattering mechanism on solid
theoretical grounds for individual systems, with arbitrary boundaries. This will
be done for dielectric waveguides, but we will show that our general results are
valid for perfect electric conductors as
well.

We assume a waveguide as shown in \fig{setup}.
The waveguide has a width $d$ whose boundaries are given by a normalized boundary function
$q(z)$, such that the boundaries are at $x=\pm d/2 \pm \sigma q$, where
$\sigma^2$ is the variance of the boundary. For a sinusoidal boundary $\sigma$ is connected to the amplitude
$A$ of the oscillation of the  boundary by $\sigma=A/\sqrt{2}$.  We chose the coordinate
transformation as
\begin{align}
		  (x,y,z) &\rightarrow(\frac{w(\tilde z)}{d}\tilde x,\tilde y, \tilde
		  z)&\textrm{with } w(\tilde z)=d+2\sigma q(\tilde z) \label{eq:trafo}
\end{align}
It will flatten the boundaries of the dielectric waveguide at $\tilde x=\pm d/2$ and thus
set $\Delta \epsilon=0$ in \eq{diel_weq_perturb}, yielding 
\begin{align}
        &\left[ \widetilde{\nabla}^2+\omega^2\mu\tilde \epsilon_0(\tilde
                x,\tilde y)  \right] \vec E(\tilde
					 x,\tilde y,\tilde z) = 0
\end{align}
The transformed Laplacian $\widetilde{\nabla}^2$ consists of several new terms
\begin{align}
	        \widetilde{\nabla}^2 &= \widetilde{\nabla}_\textrm{red}^2 +
						\tilde \nabla_b^2+
	               \tilde \nabla_x^2+
						\tilde \nabla_{sg}^2 
						\label{eq:terms}
\intertext{which are calculated in \hyperref[app:trafo]{Appendix A}. Here, the
reduced Laplacian is defined as}
\widetilde{\nabla}_\textrm{red}^2&:=\spdds{\tilde x}+\spdds{\tilde
y}+\spdds{\tilde z}\label{eq:reducedlaplacian}
\end{align}
The other terms will now be interpreted as the new dielectric
perturbation $\Delta \tilde{\epsilon}$
\begin{align}
						\tilde \nabla_b^2+
	               \tilde \nabla_x^2+
						\tilde \nabla_{sg}^2 =:
						\omega^2\mu\Delta\tilde\epsilon\label{eq:newperturb}
\end{align}
yielding the transformed wave equation
\begin{align}
        \left[ \widetilde{\nabla}_\textrm{red}^2\right. &+\left.\omega^2\mu\left( \tilde \epsilon_0(\tilde
                x,\tilde y) + \Delta\tilde\epsilon \right) \right] \vec E(\tilde
					 x,\tilde y,\tilde z) = 0\label{eq:transformed_weq}
\end{align}
In the next three sections the three terms in \eq{terms} will be studied in detail. It will
be shown that $\tilde \nabla^2_b$ yields the well-known Bragg reflection
(therefore index $b$). It is
analog to previous stratified approximations. Frequency analysis shows that $\tilde \nabla^2_x$ can be
safely neglected. Finally the last term, $\tilde \nabla_{sg}^2$ will turn out
to represent the novel mechanism of square gradient Bragg reflection (index $sg$).


\section{Stratified approximation yields Bragg scattering}
The first term
\begin{align}
			\tilde \nabla_b^2&=\left(\frac{d^2}{w^2}-1\right) \spdds{\tilde x}
\end{align}
contains no derivative of $q(z)$. In this term the curvature of the boundary has no
influence. The physical effects expected from $\tilde \nabla_b^2$
are those of the stratified waveguide.
To work out the physical influence of $\tilde \nabla^2_b$ we
ignore
the two other terms
in \eq{newperturb} and set
\begin{equation} 
        \Delta \tilde \epsilon =\frac{1}{\omega^2\mu}\left(\frac{d^2}{w^2}-1\right)\spdds{\tilde x}  
		  \label{eq:epsstrat}
\end{equation} 
The coupling coefficient from \eq{coupling_coefficient} becomes 
\begin{equation}
         C^{(b)}_{kl} =\frac{\omega}{u} \int \int d\tilde x d\tilde y   \frac{w}{d}
        \vec{\tilde{E_k}} \Delta\tilde\epsilon \vec{\tilde{E_l}}
		  \label{eq:ckl}
\end{equation}
where the prefactor is the Jacobian $dxdy=d\tilde x d\tilde y\left|
\frac{w}{d}\right|$, which is strictly positive, so that $|\frac{w}{d}|=\frac{w}{d}$.
		 The modes of the electric field $\vec{\tilde{E_k}},\vec{\tilde{E_l}}$, are the undisturbed
modes of the transformed system, i.e., $\Delta \tilde \epsilon = 0$ in
\eq{transformed_weq}. These modes are calculated in
\hyperref[app:wf]{Appendix B}.

Plugging \eq{epsstrat} into  \eq{ckl}, yields%
\begin{align}
		   C^{(b)}_{kl}
		  &=\left(\frac{d}{w}-\frac{w}{d}\right)\frac{1}{p\omega\mu}\int \int d\tilde x d\tilde y \vec{\tilde{E_k}} \spdds{\tilde x} \vec{\tilde{E_l}}
 \end{align}
			Approximating 
			$$\left( \frac{d}{w}-\frac{w}{d}
			\right)=-4\frac{\sigma}{d}q(z)\left( \frac{d+\sigma q}{d+2\sigma q} \right)\approx
			-4\frac{\sigma}{d}q(z)$$
			we have
			\begin{align}
C^{(b)}_{kl}&=q(z)I_{kl}^{(b)}\\
\intertext{where}
&I^{(b)}_{kl}= -\frac{\sigma}{d}
\frac{4}{p\omega\mu}\int \int d\tilde x d\tilde y \vec{\tilde{E_k}}
\spdds{\tilde x} \vec{\tilde{E_l}} \label{eq:ikl}
\end{align}

The coupling coefficient can now be readily calculated for given material
parameters.
Before doing so, we show, that the first term in \eq{newperturb} indeed
corresponds to Bragg scattering. 
To this end, we analyze the periodicity of $ C^{(b)}_{kl}$, which is
clearly the same as the periodicity of the corrugated boundary $q(z)$. Assuming that
$q(z)$ is a periodic function it can be expanded as 
\begin{equation}
		  q(z)=\sum_m q^{(b)}_m\, e^{-im\frac{2\pi}{\Lambda}\tilde z}
		  \label{eq:expand}
\end{equation}
We can thus write
\begin{align}
		  C^{(b)}_{kl} &=\sum_m q^{(b)}_m\, I^{(b)}_{kl} e^{-im\frac{2\pi}{\Lambda}\tilde z}\\
\end{align}
and set $\hat C^{(b)}_{kl}=q^{(b)}_mI^{(b)}_{kl}$.

As in \cite{yariv_optical_2003}, we now analyze small changes in the amplitude
$A(z)$ in \eq{yariv_6.4-12} by integrating over a length $s$, which
is long compared to $\Lambda$

\begin{equation}
        dA_k \sim \sum_l \sum_m \int_s \hat C^{(b)}_{kl}A_le^{i(\beta_k-\beta_l
					 -m\frac{2\pi}{\Lambda})z}\,dz
        \label{eq:yariv_6.4-12XXX}
\end{equation}
This integral will vanish unless the Bragg condition
\begin{equation}
\beta_k-\beta_l = m\frac{2\pi}{\Lambda}
\end{equation}
is satisfied. In case of backscattering into the same mode
($\beta_k=-\beta_l$)the Bragg
condition takes the form
\begin{equation}
		  \beta_k = m\frac{\pi}{\Lambda}
\label{eq:bragg}
\end{equation}
which is \eq{generalbragg} for lattice spacing $d=\Lambda$.

We have thus shown, that the first term of the transformation
yields the well known Bragg scattering. What about the different order $m$? For
a sinusoidal boundary $q(z)\sim \sin(\frac{2\pi}{\Lambda} z)$, there are only two
possible values $m=\pm1$ in \eq{expand}, which yield the same wavelength. 
Consequently, there is only one
single reflection resonance predicted in the stratified approximation.

After we
have successfully recovered the results for the stratified approximation we will
now turn to the next term.

\section{Beyond Bragg scattering}
%
The second term $\tilde \nabla^2_x$, comprises derivatives of $q(z)$. 
\begin{align*}
			\tilde \nabla_x^2&=
        		-\frac{\sigma}{w}\left( \pdd{q}{\tilde z}\tilde x
				\left(\spdd{\tilde x}\spdd{\tilde z} +\spdd{\tilde z} \spdd{\tilde
				x}\right)\right. \\
	  			& +\left. \pdds{q}{\tilde z}\tilde x\spdd{\tilde x}\right)
\end{align*}
Here the
curvature of the boundary influences the scattering. This is the first
indication that
we no longer deal with a stratified system. However terms in $\nabla^2_x$
will have the same periodicity as $q(z)$ itself and 
come into play at frequencies given by the Bragg condition. This means that they
act at the exact same frequencies as the terms used in the stratified
approximation. %
We will neglect this term here, because in
this study we are interested in reflection resonance that take place at
frequencies other than the Bragg resonances.

In contrast, the third term on the left hand side of \eq{newperturb} contains terms with the square of the derivative
of $q(z)$. In general  the square of a function can have a different
periodicity than $q(z)$. Therefore
we expect this term to play a role at frequencies different
from that given by the Bragg condition:
\begin{align}
		  \tilde \nabla_{sg}^2&= \frac{\sigma^2}{w^2} \left(\pdd{q}{\tilde z}\right)^2
		  		\left( 3\tilde x\spdd{\tilde x} + \tilde x^2\spdds{\tilde x}\right)
\end{align}
which yields
\begin{align}
		  C^{(sg)}_{kl} &=\left(  \pdd{q}{\tilde
			 z}\right)^2 I^{(sg)}_{kl}\label{eq:cklcurved}\\
			 \intertext{where, the overlap integral is obtained after approximating
						$w=d$ (see \eq{trafo}), as}
		  I^{(sg)}_{kl} &\approx
          \frac{\sigma^2}{d^2} \frac{1}{u\omega\mu}
                  \int d\tilde x d\tilde y 
                  \vec{\tilde{E_k}} \left(3\tilde x\spdd{\tilde x} + \tilde x^2 \spdds{\tilde x}\right)
						\vec{\tilde{E_l}}\label{eq:iklcurved}
\end{align}

Following the arguments yielding the Bragg condition
\eq{bragg}, we can expand $$\left(\pdd{q}{\tilde z}\right)^2=\sum_{m'}
p^{(sg)}_{m'}\,
e^{-im'\frac{2\pi}{\Lambda}\tilde z}$$ and derive a \textit{square gradient Bragg condition}
\begin{equation}
		  \beta_k-\beta_l = m'\frac{2\pi}{\Lambda}
\end{equation}
For backscattering into the same mode ($\beta_k=-\beta_l$) it reads
\begin{equation}
		  \beta_k= m'\frac{\pi}{\Lambda}
			\label{eq:curvedbragg}
\end{equation}
This looks just like the ordinary Bragg condition, but note, that here we have
expanded the square gradient. To see the difference we consider a specific boundary $q(z)\sim\sin(\frac{2\pi}{\Lambda}z)$. 
The Fourier series of the square gradient of $q(z)$ reads
\begin{equation}
		  \left( \pdd{\sin(\frac{2\pi}{\Lambda}z)}{z}
		  \right)^2=\left( \frac{2\pi}{\Lambda} \right)^2\left(
		  \frac{1}{2}+\frac{1}{4}e^{-i\frac{4\pi}{\Lambda}
z}+\frac{1}{4}e^{i\frac{4\pi}{\Lambda}z}
		  \right)
		  \label{}
\end{equation}
In contrast to the Bragg condition ($m=\pm 1$), here we have two contributions $m'=0$ and
$m'=2$.
This means, that the square gradient Bragg scattering impacts the transmission at two disjunct frequencies. 

Plugging $m=0$ into the square gradient Bragg condition \eq{curvedbragg}, shows that backscattering into the same mode ($\beta_k = -\beta_l$) occurs for $\beta_k\rightarrow 0$. 
In symmetric dielectric waveguides (as the exemplary waveguide, we discuss) this effect is not present because modes have
a non-zero cut-off frequency. This means, that for small wave vectors there is no
guided mode that could be affected by the $m=0$ scattering. For asymmetric
dielectric waveguides it should in principle be possible to observe strong
backscattering due to square gradient Bragg reflection for $\beta \rightarrow
0$.
This effect has been demonstrated experimentally within the statistical
approach for perfectly conducting metallic waveguides the do not have
cut-off frequencies \cite{dietz_surface_2012}.
We will show that our theoretical framework covers these results as well.
However, contrast to the statistical approach \cite{izrailev_manifestation_2006},
which works only for boundaries that feature peculiar statistic features our
approach covers arbitrary shaped boundaries. In particular, we are not longer
restricted to (pseudo) random boundaries.

The second contribution is $m=2$. It operates at half the wavelength of
ordinary Bragg scattering. This is in fact a general feature. Every boundary that obeys
$$q(z+\Lambda/2)=-q(z)$$
will exhibit square gradient Bragg resonances at half the wavelenght of the
Bragg resonance.
This is, because the square of such a $q$ will have a periodicity $\Lambda/2$. 

In contrast to Bragg scattering
the frequency domain where square gradient Bragg scattering occurs didn't
attract much attention, presumably because there was no further Bragg order
expected or the square gradient Bragg resonance was confused with higher order
Bragg resonances. Therefore, this reflection resonance has -- to our knowledge
-- not been observed or identified experimentally.

\section{Coupled mode equations for arbitrary boundaries}
So far we have qualitatively investigated infinite periodic systems, that could
be expanded in a Fourier series.  In this sections it is shown how
arbitrary (finite) boundary profiles can be treated quantitatively. We will see that the
Bragg condition will be replaced by its continuous counter-part, the
Fourier transform of the boundary. So, instead of expanding the boundary as a
Fourier series, it will now be represented as a Fourier transformation. This means dropping the assumption
of a periodic function.
\begin{align}
        q(\tilde z) &= \sum_{m}q^{(b)}_m e^{-im \frac{2\pi}{\Lambda}z} \rightarrow
        \int d\beta q_b(\beta) e^{-i \beta z} 
        \label{fouriertransform}
\end{align}
To obtain the Bragg condition in the continuous case, the $dA_k$ is, as in
\eq{yariv_6.4-12XXX}, integrated over a small domain $s$:  
\begin{align}
        dA_k &=-i\frac{\beta_k}{|\beta_k|}\sum_l \int d\beta q_b(\beta) 
        \underbrace{\int_s dz\, I^{(b)}_{kl}A_le^{i(\beta_k-\beta_l -
		  \beta)z}}_{\rightarrow 0,\,\forall\, \beta_k-\beta_l \neq \beta} 
		  \intertext{The exponential function in the second integral oscillates and will thus be
zero, as long as $ \beta_k-\beta_l \neq \beta$. Therefore, the result can be
approximated by a normalized $\delta$-Function}
        dA_k&\approx -i\frac{\beta_k}{|\beta_k|} \sum_l \int d\beta q_b(\beta) N\delta(\beta_k-\beta_l - \beta) \int_s dz\, I^{(b)}_{kl}A_l\\
        &=-i\frac{\beta_k}{|\beta_k|} \sum_l N q_b(\beta_k-\beta_l) \int_s dz\, I^{(b)}_{kl}A_l \label{eq:dAkCurved}\\
		  \intertext{and in case of backscattering, where $\beta_k = -\beta_l$}
		  &=-i\frac{\beta_k}{|\beta_k|}\sum_l N q_b(2\beta_k) \int_s dz\, I^{(b)}_{kl}A_l
\end{align}
The normalization $N$ is discussed below.
The coupling is thus proportional to $q_b(2\beta_k)$. This line of reasoning
holds for both $q_b(\beta)$ (Bragg scattering) and $q_{sg}(\beta)$ (square
gradient Bragg scattering). Instead of discrete conditions \eq{bragg}
and \eq{curvedbragg} we now have a continuous range
where scattering can occur. This continuous range is given by the
Fourier transformations $q_{b/sg}$ of the boundary and the square of the
curvature of the boundary.
Taking the derivative of \eq{dAkCurved} yields a generalized coupled mode
equation for arbitrary boundaries (compare\eq{yariv_6.4-12})

\begin{align}
		  \dd{A_k}{z} &=-i\frac{\beta_k}{|\beta_k|}N \sum_l q_b(\beta_k-\beta_l) I^{(b)}_{kl}A_l
		  \label{eq:general_cm}
\end{align}
As before, investigating two modes, yields a set of two coupled equations
($k=1,l=2$ and $k=2,l=1$). Solving this set for contra-directional
coupling ($\beta_1=-\beta_2$), yields two solution, one for the incoming mode
$A_1$ and one for the backscattered mode $A_2$.
As before, the reflectivity is given by
the ratio of the incoming and the backscattered mode. Plugging in the solutions,
they evaluate to
\begin{align}
		  R^{(b)}_{kl}&=\left|\frac{A_2(0)}{A_1(0)}\right|^2=|\tanh\left(N q_b(\beta_1
		  - \beta_2) 
		  I^{(b)}_{kl} L\right)|^2\label{eq:refl}
		  \\
		  \intertext{for Bragg scattering. Starting from $p^{(sg)}$ and $I^{(sg)}_{kl}$ instead we
arrive at the result for square gradient Bragg scattering}
R^{(sg)}_{kl}&=\left|\frac{A_2(0)}{A_1(0)}\right|^2=|\tanh\left(N q_{sg}(\beta_1
		  - \beta_2)
		  I^{(sg)}_{kl} L\right)|^2\label{eq:reflcurved}
\end{align}
This is the main result of this paper, and shall be discussed in
detail.
At first, it is in perfect agreement with the
maximum reflectivity in the periodic case, \eq{maxrefl}. By comparing \eq{maxrefl} and
\eq{refl} we can
fix the normalization $N$. The result of the two approaches has to be the same
when evaluating the coupling of two modes in an infinite sample over a range
$L$ or in a finite system of length $L$. Thus setting $N
q_b(\beta_b)=p_{m=1}$, yields $N =\frac{2\pi}{L}$.
%

At this point we are able to show, that our results include the previous results from the
statistical approach
\cite{izrailev_manifestation_2006}.
When assuming a perfectly electric conducting hollow waveguide 
$q_{b} I^{(q)}_{kl}$ and 
$q_{sg} I^{(sg)}_{kl}$ 
evaluate to (see \hyperref[app:pec]{Appendix D})
\begin{align}
		 \left(q_b I^{(b)}_{kl} \right)^2 &= \frac{1}{L_n^{(b),(AS)}}\\
		 \left(q_{sg} I^{(sg)}_{kl} \right)^2 &= \frac{1}{L_n^{(b),(SGS)}}
		  \label{eq:loclength}
\end{align}
for even modes.  $L_n^{(b),(AS/SGS)}$ is the backscattering length derived in \cite{izrailev_manifestation_2006}. 
In contrast to \cite{izrailev_manifestation_2006}, we find different results
for odd and even modes. 
However, it seems that the authors of \cite{izrailev_manifestation_2006} were unaware that
they studied a symmetry reduced version of the system. Therefore their results are only
valid for modes with $E(z=0) = 0$, i.e., even modes.

This means that we derived the very same expression under drastically relaxed assumptions.
First, our result is valid for \textit{any} boundary and the reflectivity is
directly calculated from the boundary in a straightforward way. A cumbersome
generation procedure to generate random boundaries that comply with the
statistical requirements of the statistical approach
(see \cite{dietz_surface_2012,doppler_reflection_2014}) are not necessary.
Second, the derivation does not depend on the type of boundary. Previous
studies were bound to perfectly electric conducting boundary conditions. Therefore they could not be
applied to dielectric waveguides, where the mode crosses the boundary and
penetrates the region outside the waveguide. 

\section{Reflectivity of a dielectric Waveguide}

Now the reflectivity for a dielectric
waveguide can be calculated in a straight forward manner.  
By inserting the modes of the E-Field (see \hyperref[app:wf]{Appendix B}) into
\eq{ikl} and Eq.~ \eqref{eq:iklcurved},
$I^{(b)}_{kl}$ and  $I^{(sg)}_{kl}$ are obtained. The Fourier transformation of the boundary
$q_{b}$ and of the curvature of the boundary are calculated in the
\hyperref[app:ft]{Appendix C}. Inserting these quantities into \eq{refl}
and \eq{reflcurved} yields the reflectivity. 
As an example, we investigate the first
mode in the waveguide shown in \fig{setup}.  
In \fig{R} a) and b) the calculated reflectivity $R$ and the transmission
$T$, respectively, are displayed as a function of (vacuum) wave vector $k$. 
Two sharp peaks caused by Bragg scattering into the same mode $R^{(b)}_{11}$ and
into the next odd mode $R^{(b)}_{13}$ can be clearly identified. 
The new square gradient Bragg
reflection resonances $R^{(sg)}_{11},$ and $R^{(sg)}_{13}$ occur as expected at twice the value of $\beta$.
From this calculations it is apparent that square gradient Bragg scattering is as strong as Bragg scattering.  
Note that these square gradient peaks must not be confused with conventional higher order Bragg peaks. 
In this waveguide with sinusoidal boundaries, there are no conventional higher
order Bragg terms, since the boundary has only a single frequency
component.

\section{Comparison to numerical results}
\begin{figure}
					 \includegraphics[width=0.5\textwidth]{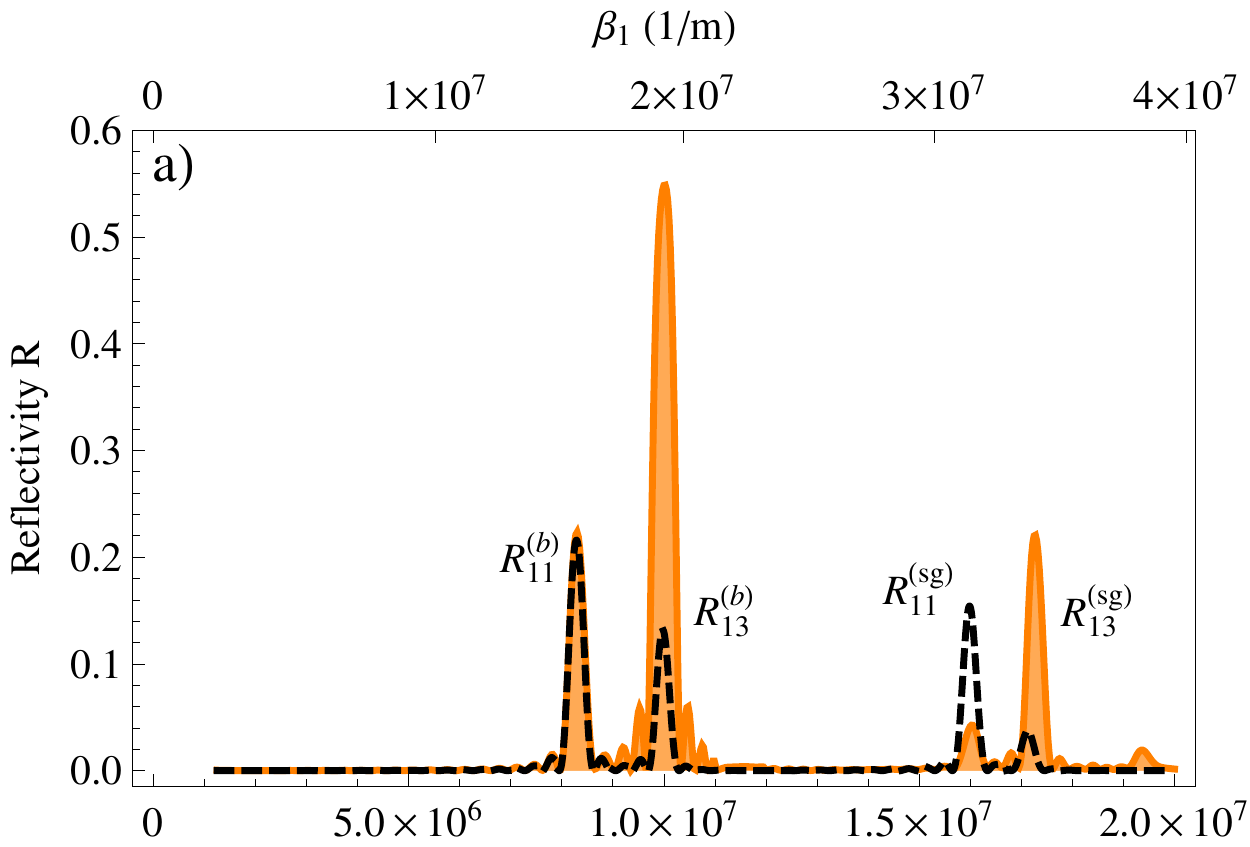}\\
					 \includegraphics[width=0.5\textwidth]{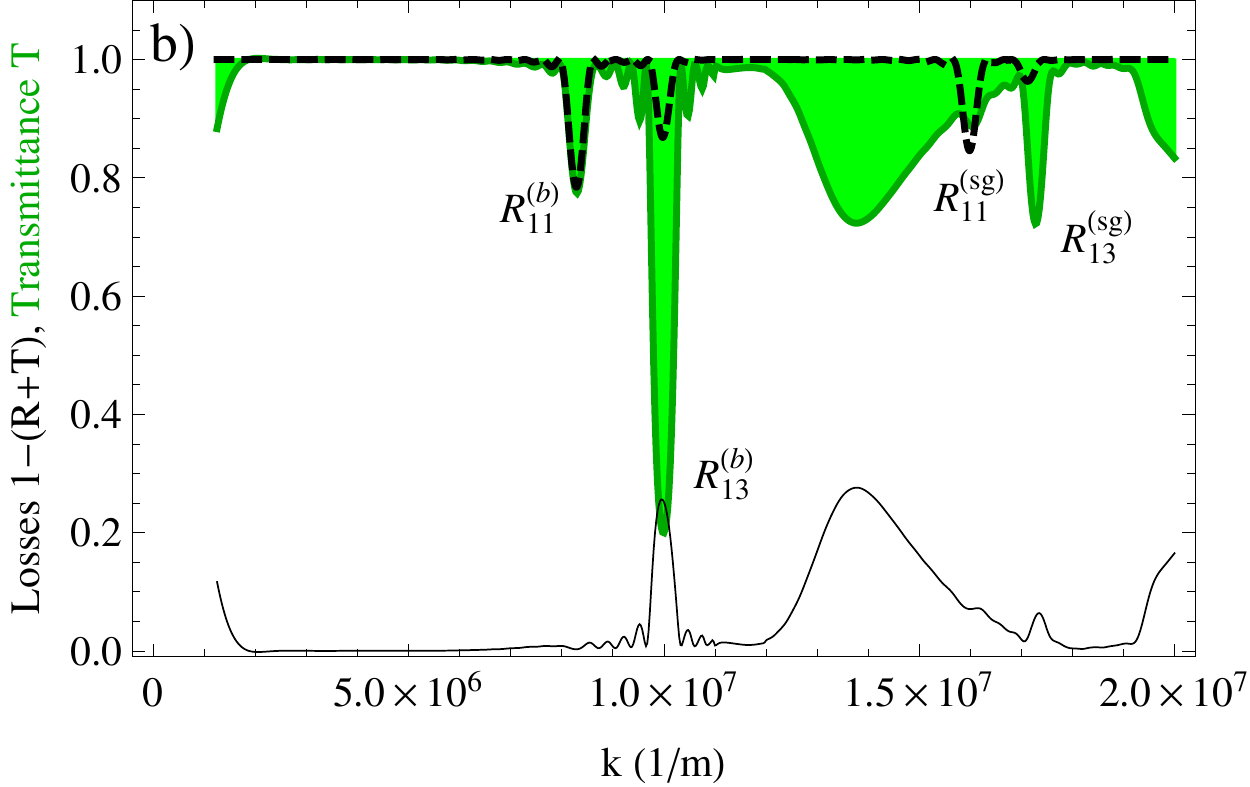}

		  \caption{a) Reflectivity of first mode through the dielectric waveguide
					 shown in \fig{setup}.
					 Analytical prediction 
					 (\AnalyticR) calculate from \eq{refl} and \eq{reflcurved} 
					 in comparison to numerical data (\Rstyle).
		  			b) Analytical transmittance calculated from reflectivity in a) as $T=1-R$ (\AnalyticT) compared to numerical
					transmission (\Tstyle). The difference in both are the radiation
					losses $1-(R+T)$ (\Lstyle).
					\label{fig:R}}
\end{figure}
To test our analytical predictions numerically we use a 
commercial finite element solver (JCMwave). 
The Maxwell equations are
solved on a non-uniform 2d mesh. 
Convergence was tested and confirmed by increasing the finite
element degree up to 7 \cite{hoffmann_comparison_2009}.
\fig{R} compares the analytical to the numerical results for the structure
displayed in \fig{setup}. 
We see in \fig{R}~ a) that  for Bragg scattering into the same mode
, $R^{(b)}_{11}$, the calculated reflectivity is in perfect agreement with the
numerical results. The situation is different for the square gradient Bragg scattering
into the same mode $R^{(sg)}_{11}$. The peak is clearly visible, but overestimated by the
theory. For inter-mode scattering $R^{(b)}_{13}$ and $R^{(sg)}_{13}$, the situation is
reversed. The reflection resonance is much stronger than predicted by the
theory, for both scattering mechanisms.  
This is surprising, because the overlap integral in $C^{(i)}_{kl}$,
\eq{ckl}, is obviously smaller for modes
where $k\neq l$ compared to $k=l$, where the overlap is maximal. 
The origin of the strongly enhanced inter-mode scattering has to be investigated
more thoroughly, in a framework beyond two-wave coupled mode theory.

To investigate the influence of square gradient Bragg scattering on the
transmission, we numerically calculated the transmission: the result 
(\Tstyle) is shown in \fig{R}~b). It is compared to the analytical transmission (\AnalyticT), calculated from the
reflectivity $T=1-R$. As expected, the reflection resonances appear as gaps in the
transmission. Still, the transmission shows strong deviation, 
from $R+T=1$, due to radiation losses.  %
\begin{figure}
		  \includegraphics[width=0.5\textwidth]{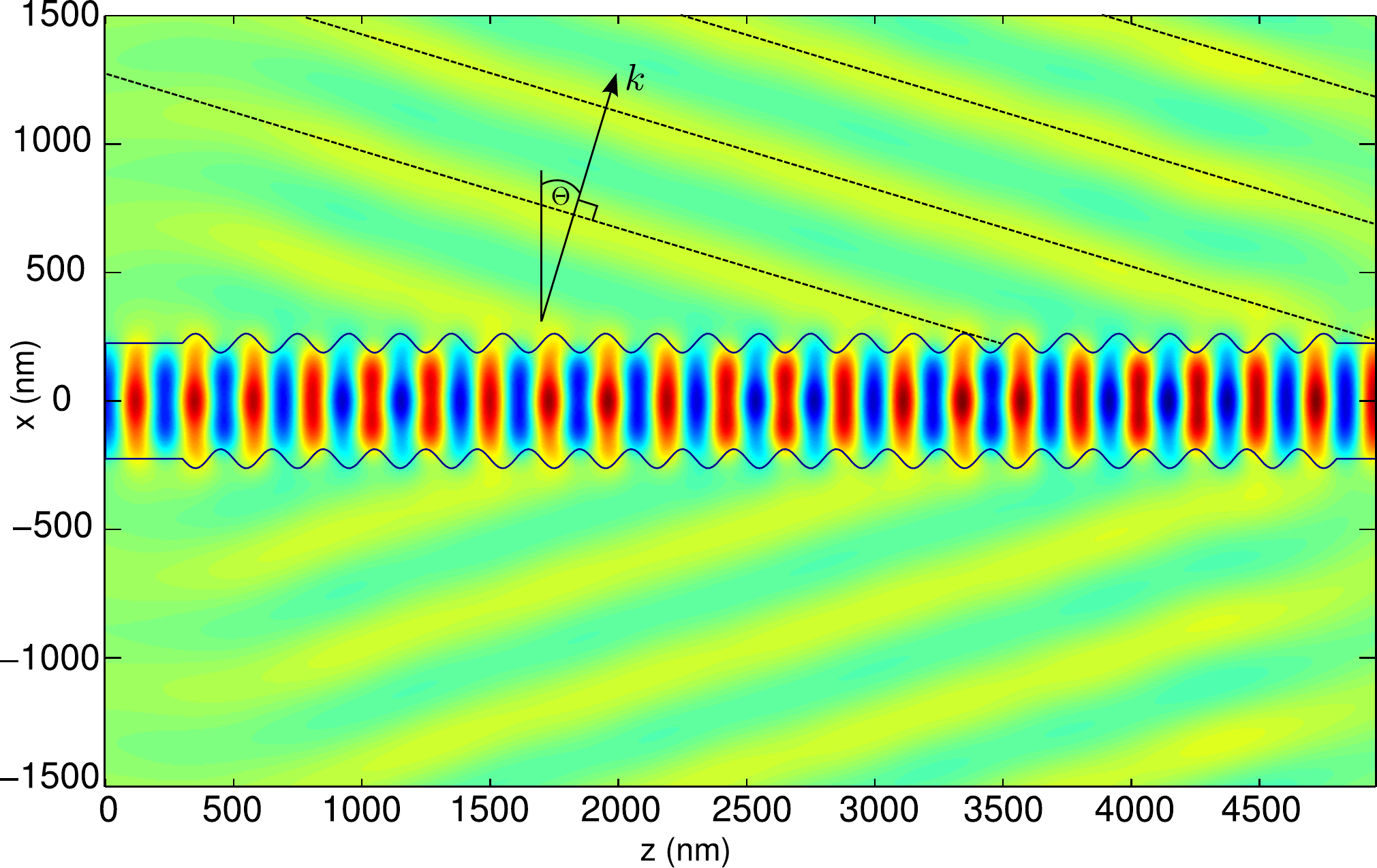}
		  \caption{E-Field ($\textrm{Re}(E_y)$) at maximum of radiation losses at
					 $k=1.4\cdot10^7/m$ (compare dip in \fig{R} b).
					 Plane waves radiate out of the waveguide
					 (dashed lines). The expected $k$-vector from \eq{coupler} is
					 indicated ($\Theta=\ang{16.78}$).
					\label{fig:radiation}}
\end{figure}

To investigate the influence of radiation losses, we plotted the losses as $1-(R+T)$
		  (\Lstyle) in \fig{R}~b).
Most apparent is the broad gap around $k = 1.4\cdot10^7/$m. 
It stems from light that is coupled out of the waveguide by coupling
to radiation modes $\beta_{out}=nk\sin\Theta$
\begin{align}
		  kn\sin\Theta + \frac{2\pi}{\Lambda}&=\beta_{k}
		  \label{eq:coupler}
\end{align}
where $n$ is the refractive index of the outer material and $\beta_{k}$ is, as
before, the propagation constant of the guided mode. 
\eq{coupler} yields (with parameters from \fig{setup}) an outcoupling angle of
$\Theta=\ang{16.78}$ (measured counter clockwise from the $x$-axis), which is in
perfect agreement with the numerically calculated maximum (\fig{radiation}).

%
The second important feature are the strong scattering losses at the
$R^{(b)}_{13}$ reflection resonance. Around 25\% of the transmission is lost due
to radiation out of the waveguide. The effect is smaller ($\sim 15\%$) for curved
Bragg scattering, $R^{(sg)}_{13}$, but still visible. 

A general observation when comparing the analytical to the numerical data is
that the reflection into the same mode is
overestimated, while backscattering into higher modes is underestimated by the generalized coupled
mode theory. The transmittance is further reduced by radiation losses at the
(square gradient) Bragg resonances.
\section{Summary}

We showed the appearance of unexpected Bragg resonances in dielectric
waveguides with corrugated boundaries. 
We generalized a
statistical approach to cover individual systems with arbitrary boundaries. 

The analytically calculated reflectivity 
is in good agreement with the numerical results. While the position of
the expected resonance is predicted with high accuracy, the strength of the
square gradient Bragg scattering is strongly underestimated. We find that Bragg and
square gradient Bragg scattering are of comparable strength. 

The transmission through the waveguide is even stronger affected by the square
gradient Bragg reflections due to radiation losses.

Since Bragg scattering is one of the key properties in corrugated waveguides.
A general theory which is able to describe all Bragg resonances 
is a promising new tool for designing optical systems based on waveguide
structures.
For example, in directional couplers, coupling and reflecting gratings are
combined \cite{kewes_key_2015}. Both gratings have to have different periodicity. Using the
square gradient Bragg scattering mechanism it should be possible to use the same grating for
coupling and reflection purposes. This would considerably simplify grating
structures. 

\section{Acknowledgments}
O.D.\,expresses his gratitude to Stefan Rotter (TU Vienna) for his encouraging critique on
the early version of this paper. 
He likes to thank Dan-Nha Huynh and Matthias Moeferdt for fruitful discussion.
Support by SFB 787 is acknowledged by O.D. 
\appendix
\section{Explicit calculation of the coordinate transformation}
\label{app:trafo}
The differential operator in the new coordinates 
\begin{align*}
		  (x,y,z) &\rightarrow(\frac{w(\tilde z)}{d}\tilde x,\tilde y, \tilde
		  z)&\textrm{with } w(\tilde z)=d+2\sigma q(\tilde z) \label{eq:trafoneu}
\end{align*}
is derived in the following.
For some function $q(\tilde x,\tilde y,\tilde z)$ we obtain
\begin{align*}
		  \spdd{x} q &= \left( \pdd{\tilde x}{x}\spdd{\tilde x} + \pdd{\tilde
		  y}{x}\spdd{\tilde y} + \pdd{\tilde z}{x}\spdd{\tilde z} \right) q \\
		  &= \left( \pdd{\tilde x}{x}\spdd{\tilde x}\right)q = 
		  \frac{d}{w} \spdd{\tilde x} q\\
		  \spdd{y} q &= \left( \pdd{\tilde x}{y}\spdd{\tilde x} + \pdd{\tilde
		  y}{y}\spdd{\tilde y} + \pdd{\tilde z}{y}\spdd{\tilde z} \right) q =
		  \spdd{\tilde y} q \\
		  \spdd{z} q &= \left( \pdd{\tilde x}{z}\spdd{\tilde x} + \pdd{\tilde
		  y}{z}\spdd{\tilde y} + \pdd{\tilde z}{z}\spdd{\tilde z} \right) q\\
		  &= \left( \pdd{\tilde x}{z}\spdd{\tilde x} + \spdd{\tilde z} \right) q\\ 
		  &= \left( \frac{d x \sigma}{-w(\tilde z)^2} \pdd{q}{\tilde
		  z}\spdd{\tilde x}+  \spdd{\tilde z}\right)q\\
		  &= \left( \underbrace{\tilde x\frac{\sigma}{-w(\tilde z)}
		  \pdd{q}{\tilde z}\spdd{\tilde x}}_A 
		  +  \underbrace{\spdd{\tilde z}}_B\right)q
\end{align*}
where $A$, and $B$, label the terms for better tracking. Applying the derivation twice yields
\begin{align*}
		  \spdds{x} q &= \frac{d^2}{w^2} \spdds{\tilde x} q\\
		  \spdds{y} q &= \spdds{\tilde y} q \\
		  \spdds{z} q &= \left( \underbrace{\tilde x\frac{\sigma^2}{w^2} \left(  \pdd{q}{\tilde z}\right)^2\spdd{\tilde x}
		  + \tilde x^2\frac{\sigma^2}{w^2} \left(  \pdd{q}{\tilde
		  z}\right)^2\spdds{\tilde x}}_{A^2}\right.\\
		  &+\underbrace{\tilde x  \frac{\sigma}{-w} \pdd{q}{\tilde z}\spdd{\tilde x}\spdd{\tilde z}}_{AB}
		  + \underbrace{\spdd{\tilde z} \tilde x \frac{\sigma}{-w(\tilde z)}
		  \pdd{q}{\tilde z}\spdd{\tilde x}}_{BA}\\
		  &+  \left.\underbrace{\spdds{\tilde z}}_{B^2} \right)q
\end{align*}
where the last term becomes
\begin{align*}
		  &\spdd{\tilde z} \tilde x \frac{\sigma}{-w(\tilde z)} \pdd{q}{\tilde z}\spdd{\tilde x}\\
		  &=\frac{\tilde x\sigma^2}{w^2} \left( \pdd{q}{\tilde z} \right)^2\spdd{\tilde x} 
		  +\frac{\tilde x\sigma^2}{w^2} \left(\pdd{q}{\tilde z}\right)^2\spdd{\tilde x} \\
		  &+\frac{\tilde x \sigma}{-w} \left(\pdds{q}{\tilde z}\right)\spdd{\tilde x} 
		  +\frac{\tilde x\sigma}{-w} \pdd{q}{\tilde z}\spdd{\tilde z} \spdd{\tilde x} 
\end{align*}
There appear different types of derivatives of $q$. Grouping these terms
using the reduced Laplacian defined in \eq{reducedlaplacian}
%
\begin{align*}
        \widetilde{\nabla}^2 &= \widetilde{\nabla}_\textrm{red}^2 +
						\tilde \nabla_b^2+
	               \tilde \nabla_x^2+
						\tilde \nabla_{sg}^2 
						\intertext{we have}
\tilde \nabla_b^2&=
			\left(\frac{d^2}{w^2}-1\right) \spdds{\tilde x} \\
\tilde \nabla_x^2&=
			  -\frac{\sigma}{w} \pdd{q}{\tilde z}\tilde x\left( \spdd{\tilde x}\spdd{\tilde z} 
		  + \spdd{\tilde z} \spdd{\tilde x} \right)
		  -\frac{ \sigma}{w} \left(\pdds{q}{\tilde z}\right)\tilde
		  x\spdd{\tilde x}\\
		  \tilde \nabla_{sg}^2&=
			\frac{\sigma^2}{w^2} \left(  \pdd{q}{\tilde z}\right)^2 3\tilde x\spdd{\tilde x}
		  + \frac{\sigma^2}{w^2} \left(  \pdd{q}{\tilde z}\right)^2 \tilde x^2\spdds{\tilde x}
\end{align*}
Note that only the $x$ coordinate was transformed, hence $\tilde z=z$ and $\tilde y=y$.

\section{Modes of the electric field in dielectric waveguides}
\label{app:wf}
\fig{radiation} shows the first mode of the electric field in the waveguide. 
The mode numbers above are chosen to be $k$ and $l$, to identify two specific,
but not necessarily different modes. In the following the general index $m$ is
used.
In the substrate ($\tilde x\leq-d/2$) and cover ($d/2\leq \tilde x$) regions, the $m$-th mode is given by the E-field
perpendicular to the $x$-$z$-plane as (see \cite{kogelnik_2._1975})
\begin{align*}
		  \tilde E_m(\tilde x, k)&= \begin{cases}
					 E_s(k)\exp\left(\gamma_s \left(\tilde x +
					 \frac{d}{2}\right)\right)& \textrm{in substrate} \\
E_c(k)\exp\left(-\gamma_c \left(\tilde x - \frac{d}{2}\right)\right)& \textrm{in cover} \\
				\end{cases}
				\intertext{and in the film region ($-d/2 < \tilde x  < d/2$) odd and even modes are given by}
		  \tilde E_m(\tilde x, k)&= E_f(k)\begin{cases}
					 \cos\left(\kappa_f \tilde x\right)& \textrm{for odd modes}\\
		  \sin\left(\kappa_f \tilde x\right)& \textrm{for even modes}
				\end{cases}
\end{align*}
with 
\begin{align*}
		  \kappa_c^2 &= n_c^2k^2-\beta_m^2=-\gamma_c^2\\
		  \kappa_f^2 &= n_f^2k^2-\beta_m^2 \\
		  \kappa_s^2 &= n_s^2k^2-\beta_m^2=-\gamma_s^2
\end{align*}
Here, the refractive indices refer to the substrate ($n_s$), cover ($n_c$), and
film ($n_f$) material. The wave vector $n_{c,f,s}k$ has components in direction of
propagation $\beta_m$ and in transversal direction denoted by $\kappa_{c,f,s}$.
Note, that there is one $\beta_m$ for each mode. Consequently $\kappa$ and
$\gamma$  depend on the mode number just as $\tilde E_m$.
In the case of a symmetric waveguide the allowed $\beta_m$ for each mode $m$ can
be found by numerically solving (see \cite{kogelnik_2._1975})
\begin{equation*}
        2d \kappa_f - 4 \arctan \sqrt{
                \frac{1-\left(\frac{n_s}{n_{eff}}\right)^2}
                {\left(\frac{n_f}{n_{eff}}\right)^2-1}} 
                = 2\pi m
\end{equation*}
where $n_{eff}=\beta_m/|\vec k|$.
The total normalization is chosen as
\begin{equation*}
         \frac{2\beta_m}{u\omega\mu}\int dx E_m^2 = 1\textrm{W/m}
\end{equation*}
and yields the peak field inside the waveguide for odd/even modes
\begin{align*}
        E_f^2=\frac{u\omega\mu}{\beta_m}\Bigg(
        &d\pm\frac{\sin(d\kappa_f)}{\kappa_f}\\ 
        &+\frac{n_f^2-n_{eff}^2}{n_f^2-n_c^2}\frac{1}{\gamma_{s}}
        +\frac{n_f^2-n_{eff}^2}{n_f^2-n_s^2}\frac{1}{\gamma_{c}}\Bigg)^{-1}
        \textrm{W/m}
\end{align*}
The amplitudes are connected via
\begin{align*}
        E_f^2(n_f^2-n_{eff}^2)=E_s^2(n_f^2-n_s^2)=E_c^2(n_f^2-n_c^2)
\end{align*}

\section{Fourier Transform $q_{b}$ and $q_{sg}$ of the boundary}
\label{app:ft}
\newcommand {\sinc}[1]{\textrm{sinc}\left(#1\right)}
\newcommand {\rect}{\textrm{box}^L_0(z)}
\newcommand {\rectcenter}{\textrm{box}^{L/2}_{-L/2}(z)}

Let the (normalized) boundary $q$ of the waveguide (see \fig{setup}) be of the simplest possible form:
\begin{equation*}
		  q(z) = \sqrt{2}\sin(\beta_bz)\rect
\end{equation*}
where $\beta_b=\frac{2\pi}{\Lambda}$ is the periodicity of the boundary and $\rect$ is a box, or rectangular, function of length $L$ constructed via the Heaviside
function $\Theta$, as $\rect=\Theta(z)\left(1-\Theta(z-L)\right)$.
The wave vector of the boundary roughness $\beta_b$ should be chosen in such a way
that $\beta_bL$ are multiples of $2\pi$ to ensure a continuous function. 
The Fourier transform is normalized such that 
\begin{align*}
		  q(\beta) &= \ft{q(z)}=\frac{1}{2\pi} \int q(z) \exp \left( i \beta z \right)\\
\end{align*}

To calculate $q_b$ and $q_{sg}$ the following derivatives with respect to $z$ have to be calculated
\begin{align*}
		  q' &= \sqrt{2}\beta_b\cos(\beta_bz)\rect + \sqrt{2} \sin(\beta_bz)\left(\delta(z)-\delta(z-L)\right)\\
q'^2 &= 2 \left(  \beta_b\cos(\beta_bz)\rect\right)^2\\
&\,+ 2\Big( \sin(\beta_bz)\left(\delta(z)-\delta(z-L)\right)  \Big)^2\\
&\,+ 4 \beta_b\cos(\beta_bz)\rect
\sin(\beta_bz)\left(\delta(z)-\delta(z-L)\right)
\end{align*}
Now the Fourier transformation of $q'^2$
has to be calculated. However, the boundary profile is chosen in such a way,
that it ends at $x=0$, i.e., at the position of the unperturbed
boundary. Therefore $\sin(\beta_b L)=0$, and thus the second and the third term vanishes since
\begin{align*}
 \int &\sin^2(\beta_bz)\left(\delta^2(z)-\delta^2\left(z-L\right)\right)\exp(-ikz)\\
& = \sin^2(0)\exp(0) - \sin^2(\beta_bL)\exp(-ikL) \\
& = 0
\end{align*}
Using $\rect^2=\rect$ and rewriting the square of the cosine as sum of
two cosines, yields
\begin{align*}
\ft{q'^2} &=2 \beta_b^2\ft{\cos(\beta_bz)^2\rect^2}\\
			  &=2 \beta_b^2\ft{\left(\cos(0)+\cos(2\beta_bz)\right)\rect}\\
			  &=2 \beta_b^2\ft{\rect} + \beta_b^2\ft{\cos(2\beta_bz)\rect}
\end{align*}
The intermediate result in Fourier representation is
\begin{align}
\ft{q} &= \sqrt{2}\ft{\sin(\beta_bz)\rect}\nonumber\\
\ft{q'^2} &=2 \beta_b^2\left( \ft{\rect} + \ft{\cos(2\beta_bz)\rect} \right)
\label{eq:intermediate}
\end{align}
With the convolution theorem 
\begin{equation*}
\ft{fg} = \ft{f}*\ft{g} 
\end{equation*}this result can be further simplified. The ($*$) operator denotes a
convolution. Especially interesting for the present case is the convolution with
a $\delta$-function, which evaluates as
\begin{equation*}
f(z)*\delta(z-b) = f(z-b).
\end{equation*}
Shifting a function, will add an additional phase factor to its Fourier
transformation. Therefore instead of evaluating the functions in
\eq{intermediate} directly they will be shifted to be centered around
$z=0$.  The Fourier transformation of a centered box function is given by 
\begin{equation*}
		  \ft{\rectcenter} = \frac{L}{2\pi}\,\sinc{\frac{\beta L}{2}}
\end{equation*}
and the Fourier transformation of the Cosine and Sine is given by
\begin{align*}
		  \ft{\sin{\beta_bz}} &= \frac{i}{2} \delta(\beta - \beta_b) - \frac{i}{2}  \delta(\beta + \beta_b)\\
\ft{\cos{2\beta_bz}} &= \frac{1}{2} \delta(\beta - 2\beta_b) + \frac{1}{2}\delta(\beta + 2\beta_b)
\end{align*}
Applying the convolution theorem and evaluating the $\delta$-functions the
resulting expression reads
\begin{align*}
		  \ft{q}&= \frac{L}{2\pi}
		  \frac{i}{\sqrt{2}}\left(\sinc{\frac{1}{2}(\beta-\beta_b) L} \right.\\
		  &\left.- \sinc{\frac{1}{2}(\beta+\beta_b) L} \right)\\
		  \ft{q'^2} &= \frac{L}{\pi} \beta_b^2\,\left(\sinc{\frac{\beta L}{2}} +
		  \frac{1}{2}\sinc{\frac{1}{2}(\beta-2\beta_b) L}\right) \\
		  &+\frac{L}{\pi} \beta_b^2\, \left(\frac{1}{2} \sinc{\frac{1}{2}(\beta+2\beta_b) L} \right)
\end{align*}
Now one can neglect the last terms in both equation*, because they contribute for
negative frequencies only. 
The expression for $q_{sg}$ can be separated in two different contributions.
The first contribution, responsible for a peak at $\beta=0$ is nothing but the
Fourier transformation of the box-function. The individual shape of the boundary
has no influence for small $\beta$.  
The second term is responsible for a peak at $\beta=2 \beta_bx$, similar to the
		  peak of $q_{b}$ at $\beta=\beta_bz$.

\section{hollow waveguide with perfectly conducting walls}
\label{app:pec}
In this section we will now apply
our findings to special case of hollow waveguides, with perfectly reflecting boundaries. 
We will show, that previous theoretical work is included in our theory.

Assume a hollow metallic waveguide, such as a microwave waveguide. It is convenient to assume a perfect electric conductor. That is assuming that the E-Field vanishes at the boundaries. The longitudinal wave vector $\beta$ is than a simple function of $k$
\begin{equation*}
         \beta = \sqrt{|\vec k|^2 - \left(\frac{\pi n}{d}\right)^2}.
\end{equation*}
$\beta$ can take any number from 0 to $\infty$. This means that $\vec k$ can have
any orientation from transversal to nearly longitudinal. The mode will
be zero everywhere except inside the waveguide, between $x=-d/2$ and
$x=d/2$. This means that the dielectric mode given in 
\hyperref[app:wf]{Appendix B}, is
drastically simplified by $\gamma_c=\gamma_s=0$ and
$\kappa_f=\frac{\pi n}{d}$. It is thus restricted to interior of the waveguide
$-d/2<x<d/2$:
\begin{align*}
		  \tilde E_y(x, k)&= E_o(k)\begin{cases}
					 \cos\left( \frac{\pi n}{d}\tilde x\right)& \textrm{for odd modes}\\
		  \sin\left(\frac{\pi n}{d}\tilde x\right)& \textrm{for even modes}
				\end{cases}
\end{align*}
Normalizing the power to $1$ W/m yields $E^2_0=\frac{p\omega \mu}{\beta d}$.

Now $I^{(b)}_{kl}$ and $I^{(sg)}_{kl}$ can be calculated as
\begin{align*}
		  \hat C^{(b)}_{kl}&= q_b(2\beta) I^{(b)}_{kl}=\frac{\sigma}{d^3}
            \frac{2 \pi^2n^2}{\beta} q_b(2\beta)  \\\label{eq:couplingpec}
        {C^{(sg)}_{kl}}&=q_{sg}(2\beta) I^{(sg)}_{kl}=\frac{1}{2}\frac{\sigma^2}{d^2\beta}
		  \left(1- \frac{\pi^2 n^2}{12} \right)q_{sg}(2\beta)& \textrm{for odd $n$}\\
        {C^{(sg)}_{kl}}&=q_{sg}(2\beta) I^{(sg)}_{kl}=\frac{1}{2}\frac{\sigma^2}{d^2\beta}
		  \left(1+ \frac{\pi^2 n^2}{3}
        \right)q_{sg}(2\beta)& \textrm{for even $n$}
\end{align*}
After identifying $|q_b(2\beta)|^2$ with $W(2\beta)$ and  $|\hat
p_{sg}(2\beta)|^2$ with $2 S(2\beta)$ \cite{dietz_surface_2012}, we see a
surprisingly simple relationship between coupling coefficient and localization
length $L_n$ for the even modes:
\begin{align*}
		  \frac{1}{L_{n}^{(b),(AS)}}&=(\hat C^{(b)}_{kl})^2\\
		  \frac{1}{L_{n}^{(b),(SGS)}}&=(C^{(sg)}_{kl})^2
\end{align*}

\bibliography{biblio}

\end{document}